\documentclass[twocolumn,a4paper,10pt]{article}
\usepackage{amsmath,amsfonts}
\usepackage{epsfig}
\usepackage{times}
\usepackage{epsfig}
\usepackage[small,hang]{caption}
\usepackage{siunitx}
\DeclareMathOperator{\sign}{sign}
\usepackage{amssymb}

\makeatletter  

\newcounter{numbersec}
\renewcommand{\section}[1]{\par\noindent\stepcounter{numbersec}
	\par
	\vspace{6pt}
	\noindent\textbf{\large   \arabic{numbersec} \hspace*{0.3cm} #1 }
	\par
	\vspace{2pt}
}
\renewcommand{\subsection}[1]{
	\par
	\vspace{6pt}
	\noindent\textbf{#1}
	\par
}
\renewcommand{\subsubsection}[1]{%
	\par
	\vspace{6pt}
	\textbf{#1.}
}

\newcommand{\Abstract}{\par\vspace{6pt}\noindent\textbf{\large Abstract}\par\vspace{2pt}}
\newcommand{\Acknowledgments}{\par\vspace{6pt}\noindent\textbf{\large Acknowledgments }\par\vspace{2pt}}

\newenvironment{References}{
\par\vspace{6pt}\noindent\textbf{\large References}\par\vspace{2pt}
\begin{small}\begin{list}{ }{\itemsep2mm \parsep0mm\labelsep0mm\leftmargin0mm}}
{\end{list}\end{small}}

\makeatother

\title{\vspace*{-12mm}
\LARGE \sc \textbf{  
Application of riblets to separating turbulent \\
boundary layers
}}

\author{ \Large \bf \textit{ 
A.\ Rouhi$^{1}$, V.\ Kumar$^{2}$, O.\ Lehmkuhl$^{2}$, W.\ Wu$^{3}$, M.\ Kozul$^{4}$ and A. J. Smits$^{5}$ }  \\ \\
\bf  $^{1}$ \textit{ Department of Engineering, Nottingham Trent University, Nottingham, UK} \\
\bf  $^{2}$ \textit{ CASE, Barcelona Supercomputing Center (BSC), Barcelona, Spain} \\
\bf  $^{3}$ \textit{ Department of Mechanical Engineering, University of Mississippi, Oxford, USA} \\ 
\bf  $^{4}$ \textit{ Department of Mechanical Engineering, University of Melbourne, Melbourne, Australia} \\
\bf  $^{5}$ \textit{ Department of Mechanical and Aerospace Engineering, Princeton University, Princeton, USA} \\
{\it amirreza.rouhi@ntu.ac.uk}
}
\date{}

\begin{document}
\maketitle
\thispagestyle{empty}



\Abstract

 We conduct direct numerical simulations of separating turbulent boundary layers (TBLs) over triangular riblets with tip angles $90^o$ (T9) and $60^o$ (T6). Our setup follows the separating TBL study of Wu et al.\ ({\it J. Fluid Mech.}, vol.\ 883, 2020, p.\ A45). An equilibrium zero pressure-gradient (ZPG) TBL is generated at a reference location, followed by imposition of a Gaussian suction profile to create a separation bubble. The ZPG TBLs over the riblets and the benchmark smooth case have matched momentum thickness Reynolds number $Re_{\theta_0} = 583$ (friction Reynolds number 224). We employ a well-validated spectral-element solver, and leverage its unstructured-grid nature to generate an optimal grid, based on the size of turbulent scales across the TBL. At the reference location, the T9 and T6 riblets respectively increase and reduce drag, with viscous-scaled spacings $52$ and $13$. We discover that for both riblet cases, the mean separation location occurs at a distance of $140\theta_0$ downstream of the reference location, $18\%$ shorter than the mean separation distance for the smooth case ($170\theta_0$). This outcome is related to the progressive enhancement of the Kelvin-Helmholtz (KH) rollers over the riblets, owing to the continuous rise in the adverse pressure-gradient. The KH rollers penetrate into the turbulent separation bubble, with significantly larger size and coherence compared to their counterparts upstream of the mean separation location.


\section{Introduction} 

Thanks to the great advancements by the turbulence community, riblets are known to reduce turbulent skin-friction drag, and the underlying physics are unraveled to a great detail (Garcia-Mayoral et al.\ 2019). These advancements mainly focus on equilibrium turbulent flows. Most computations focus on turbulent channel flows (Choi et al.\ 1993, Garcia-Mayoral \& Jim{\'e}nez 2011, Endrikat et al.\ 2021), and most experiments focus on zero pressure-gradient (ZPG) turbulent boundary layers (TBLs) (Walsh \& Weinstein 1979, Choi 1989). However, the response of riblets to non-equilibrium flows is investigated to a less extent. In the present study, we focus on TBL separation under a strong Adverse Pressure-Gradient (APG), as a prominent non-equilibrium flow in the Aerospace industry–diffusers, wing trailing edge, and low-pressure turbines to name a few.

Several studies have investigated the response of riblets under APG TBL (Nieuwstadt et al.\ 1993, Debisschop \& Nieuwstad 1996, Klumpp et al.\ 2010,  Boomsma \& Sotiropoulos 2015). The consensus is that riblets yield a higher drag-reduction under APG than ZPG, and drag-reduction increases by increasing APG. However, these investigations consider attached TBL under mild APG. Quantitatively, they consider Clauser pressure-gradient parameter $\beta \lesssim 2.2$, where $\beta \equiv (\delta^*/\tau_w)(dP/dx)_e$, and $\delta^*, \tau_w$ and $(dP/dx)_e$ are respectively the displacement thickness, mean wall shear-stress and mean pressure-gradient at the edge of the TBL. Among the above-mentioned studies, Debisschop \& Nieuwstad (1996) applied the highest $\beta = 2.2$, and achieved the highest drag-reduction of $13\%$ over riblets. 

In a recent Direct Numerical Simulation (DNS) study, Savino et al.\ (2025) applied strong APG TBL over sinusoidal riblets. The APG TBL was attached over the benchmark smooth wall, with $0 \le \beta \le 10$. The authors reported drag reduction up to $100\%$ by riblets. Interestingly, the wide-spacing riblets that increase drag under ZPG, yielded higher drag reduction than the narrow-spacing riblets that reduce drag under ZPG. The authors discovered that the riblet-generated Kelvin-Helmholtz (KH) rollers are augmented under APG, in terms of size and frequency; the KH rollers create larger patches of flow reversal, hence reducing $\tau_w$. Here, we extend the study of Savino et al.\ (2025) to separating TBL. We impose a very strong APG to create massive TBL separation over the riblets, as well as the benchmark smooth wall. We aim to study the role of riblets in modifying the separation point, and the underlying physics.

\section{Methodology}
\subsection{Governing equations and solution method}
We solve the incompressible conservation of mass and momentum equations with constant density $\rho$ and kinematic viscosity $\nu$:
\begin{align}
 \boldsymbol{\nabla}\boldsymbol{\cdot}\mathbf{u} = 0, \quad
 \frac{\partial \mathbf{u}}{\partial t} + \boldsymbol{\nabla}\boldsymbol{\cdot}\mathbf{(uu)} = -\frac{1}{\rho}\boldsymbol{\nabla}p + \nu \nabla^2 \mathbf{u}. \tag{1\textit{a,b}} \label{eq:cont_mom}
\end{align}
In our notation, $\mathbf{u} = (u,v,w)$ is the velocity vector, and $x,y,z$ are the streamwise, wall-normal and spanwise directions, respectively. Equations (1\textit{a,b}) are solved using a well-validated computational solver (SOD2D) developed by Barcelona Supercomputing Center. Spatial discretization is a spectral formulation of the Continuous Galerkin Finite Elements model
coupled with a modified version of Guermond’s Entropy Viscosity stabilization model (Guermond et al.\ 2011). Time is advanced using the 2nd-order Adams-Bashforth scheme for the convection terms, and the 2nd-order Crank-Nicolson for the diffusion terms. Equations (\ref{eq:cont_mom}) are marched using fractional-step algorithm.

\begin{figure}
	\begin{center}
	\includegraphics*[width=\linewidth]{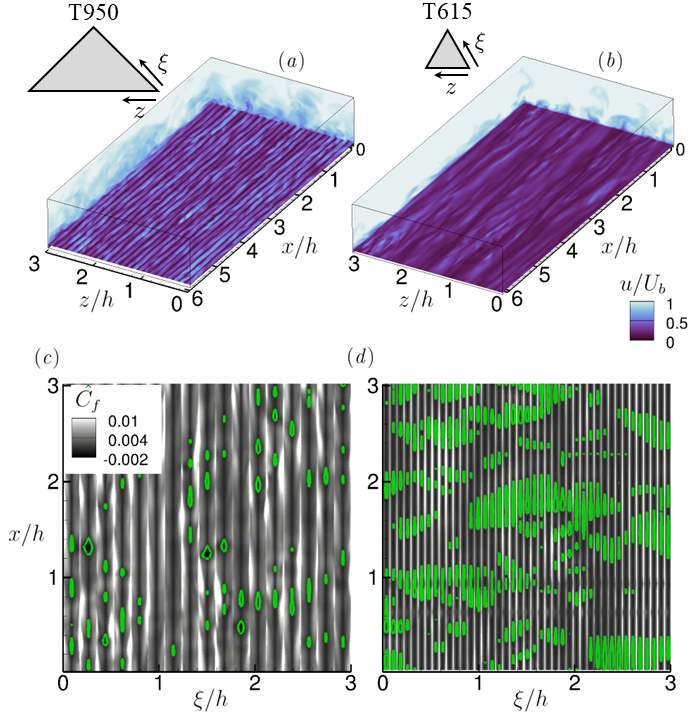}
	\caption{\label{fig:fig1} Channel flow setup at $Re_\tau = 400$ over (\textit{a,c}) T950, and (\textit{b,d}) T615 riblets (from Table~\ref{tab:channel}). (\textit{a,b}) visualize the instantaneous streamwise velocity $u$ near the riblet crest and on the side boundaries. (\textit{c,d}) visualize the instantaneous $\hat{C}_f$ over the streamwise ($x$) and azimuthal ($\xi$) coordinates. Contours of $\hat{C}_f = 0$ are highlighted in green.}
	\end{center}
\end{figure}

\subsection{Riblet geometries}
We focus on two isosceles triangular riblet geometries (Figure~\ref{fig:fig1}), termed T950 and T615, following the naming convention by Endrikat et al.\ (2021). Riblet T950 has tip angle $\alpha = \ang{90}$ and viscous-scaled spacing $s^+ = 50$, and T615 has $\alpha = \ang{60}$ and $s^+ = 14.7$. In a turbulent channel flow at friction Reynolds number $Re_\tau = 400$, T950 and T615 respectively increase and reduce drag. We assess the accuracy of SOD2D and its grid resolution requirements for these riblet geometries by replicating the DNSs of Endrikat et al.\ (2021) in a turbulent half-channel flow (Figure~\ref{fig:fig1}). For each riblet geometry we consider two grid resolutions (noted as ``coarse'' and ``fine''), with details provided in Table~\ref{tab:channel}. We generate a uniform grid in the streamwise direction, while the wall-normal grid size follows a hyperbolic tangent distribution from $\Delta^+_y \simeq 0.2 - 0.7$ at the bottom wall to $\Delta^+_y \simeq 5.0 - 8.0$ at the top boundary. We apply a uniform azimuthal grid size $\Delta^+_\xi \simeq 0.4 - 1.0$ over the riblets, and we gradually expand it to the spanwise grid size $\Delta^+_z \simeq 4.0 - 12.0$ at the top boundary. We further discuss our grid-generation approach when we present our separating TBL setup (Figure~\ref{fig:fig4}).

\begin{figure}
	\begin{center}
	\includegraphics*[width=0.8\linewidth]{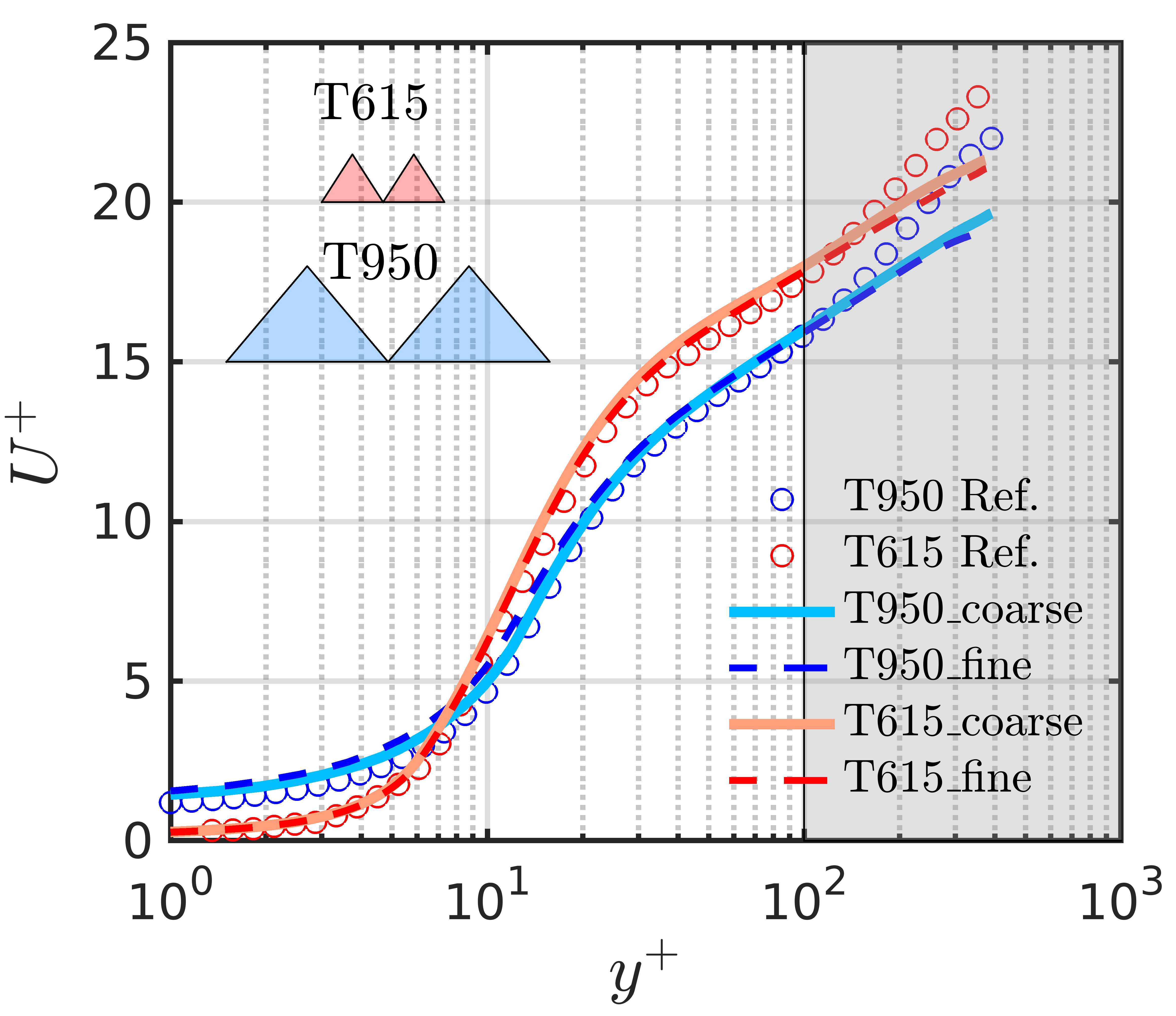}
	\caption{\label{fig:fig2} Mean velocity profiles for the cases from Table~\ref{tab:channel} compared with the reference data of Endrikat et al. (2021). The shaded region ($y^+ > 100$) is discarded, as the reference profiles are from a reduced-domain channel flow simulation.}
	\end{center}
\end{figure}

\begin{table}
\centering
\begin{tabular}{cccccc}
 Case  & $ \Delta^+_x$ & $\Delta^+_\xi$ & $\Delta^+_z$ & $\Delta^+_y$   \\ \hline
 T950\_{coarse}  &  $20.0$ & $1.0$ & $0.7 - 4.0$ & $0.7 - 8.1$  \\
 T950\_{fine}  &  $10.0$ & $1.0$ & $0.7 - 4.0$ & $0.7 - 8.1$ \\ \hline
 T615\_{coarse}  &  $10.0$ & $0.6$ & $0.3 - 12.0$ & $0.25 - 7.87$  \\
 T615\_{fine}  &  $7.0$ & $0.4$ & $0.2 - 8.2$ & $0.17 - 5.13$ \\ \hline
  \end{tabular}
\caption{Grid resolution details for DNSs of turbulent half-channel flow over T950 and T615 riblets; $\Delta^+_\xi$ is the viscous-scaled azimuthal grid size over the riblets.} \label{tab:channel}
\end{table}

Figure~\ref{fig:fig2} shows very good agreement in the mean velocity profiles between our setups from Table~\ref{tab:channel} and the reference DNSs of Endrikat et al.\ (2021). Furthermore, the velocity profiles from the coarse and fine cases are almost identical. This indicates the accuracy of SOD2D, even with $\Delta^+_x = 20$.

A notable difference between T950 and T615 is in the prominence of the KH rollers. In Figures~\ref{fig:fig1}(\textit{c,d}), we identify the KH rollers by plotting local skin-friction coefficient $\hat{C}_f \equiv 2\hat{\tau}_w/(\rho U^2_b)$ over the streamwise ($x$) and azimuthal ($\xi$) coordinates, where $\hat{\tau}_w = \sign(\partial u/\partial y)\sqrt{(\partial u/\partial y)^2 + (\partial u/\partial z)^2}$. Negative patches of $\hat{C}_f$ (marked by green iso-lines) highlight the local flow reversals by the KH rollers. For T615, patches of $\hat{C}_f < 0$ occupy $7.2\%$ of the wetted area, whereas for T950 they occupy $1.8\%$. The formation of the KH rollers has a strong influence on the separating TBL, as discussed in the results section.


\subsection{Separating boundary layer setup}\label{sec:sbl}
\begin{figure}
	\begin{center}
	\includegraphics*[width=\linewidth,trim={{0.0\linewidth} {0.0\linewidth} {0.0\linewidth} {0.0\linewidth}},clip]{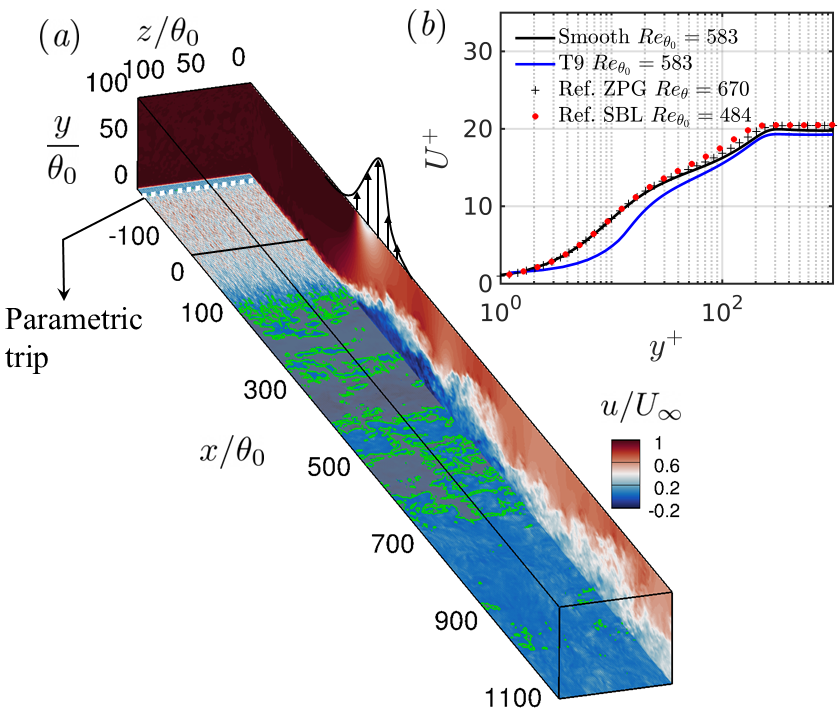}
	\caption{\label{fig:fig3} (\textit{a}) Flow setup and visualization for the T9 case. The $xz$-plane visualization is near the riblets crest with the patches of flow reversal $(u < 0)$ enveloped with green isoline. (\textit{b}) Mean velocity profiles at $x=0$ for the smooth case and the T9 case compared with the reference profiles of Schlatter \& {\"O}rl{\"u} (2010) (Ref.\ ZPG) and Wu et al.\ (2020) (Ref.\ SBL).}
	\end{center}
\end{figure}
Figure~\ref{fig:fig3}(\textit{a}) illustrates our setup for the separating TBL over the triangular riblets with $\alpha = \ang{90}$ (called T9 case). We build a similar setup for the separating TBL over the smooth wall and over the triangular riblets with $\alpha = \ang{60}$ (called T6 case), with identical domain dimensions and suction profiles. Our setup follows the DNS of separating TBL over a smooth wall by Wu et al.\ (2020). A ZPG TBL is generated at a reference location ($x=0$); at downstream the TBL is separated by applying a suction profile at the top boundary, as formulated by Wu et al.\ (2020) 
\begin{align}
 V_\mathrm{top} = V_\mathrm{max} \exp \left[ - \left( \frac{x - x_\mathrm{suc}}{\alpha L_y} \right)^2 \right] \tag{2} \label{eq:suction}
\end{align}
where $V_\mathrm{max} = 0.9 U_{\infty_0}, x_\mathrm{suc} = 171.1 \theta_0, \alpha = 0.3375, L_y = 99.6 \theta_0$, and $U_{\infty_0}$ and $\theta_0$ are respectively the free-stream velocity and momentum thickness at $x=0$. The suction profile (\ref{eq:suction}) triggers a strong APG, similar to the one over the suction surface of a typical airfoil near stall. To generate the ZPG TBL, we apply a laminar boundary layer at the inflow  with displacement thickness Reynolds number $Re_{\delta^*} = 775$, that is tripped by a parametric forcing. For the smooth case, the laminar inflow is the Blasius profile. For the riblet cases, we obtain the laminar inflow from a precursor temporal boundary layer simulation up to the target $Re_{\delta^*} = 775$. The smooth and riblet cases respond differently to the tripping, i.e.\ the resulting ZPG TBL reaches a target $Re_\theta$ at different distances downstream. For each case, we ensure that we place $x=0$ where $Re_{\theta_0} = 585$ ($Re_{\tau_0} \simeq 224$), hence all our smooth and riblet cases have identical $U_{\infty_0} = 1.0, \nu = 1/775, \theta_0 = 0.755$, with identical suction profile and location. At $x=0$, the T9 case has $s^+_0 = 52.2$, and the T6 case has $s^+_0 = 12.7$; these riblet spacings are close to the ones for T950 and T615 (Figure~\ref{fig:fig1}), that we extensively studied their grid resolution requirements (Figure~\ref{fig:fig2}). Figure~\ref{fig:fig3}(\textit{b}) shows the ZPG TBL profiles at $x=0$ for the smooth case and the T9 case. The smooth case profile is close to the reference profiles of Schlatter \& {\"O}rl{\"u} (2010) and Wu et al.\ (2020); some slight differences exist in the wake region. These are related to the differences in the inflow generation techniques and $Re_\theta$. Nevertheless, we show that such differences have marginal influence on the separating TBL (Figure~\ref{fig:fig5}). We apply periodic boundary condition in the spanwise ($z$-) direction, zero-vorticity condition on the top boundary, and convective boundary condition at the outlet. For all cases, the effective domain length from $x=0$ is $L_{x_\mathrm{eff}} \gtrsim 991 \theta_0$, which is larger than the one by Wu et al.\ (2020) ($L_{x_\mathrm{eff}} \simeq 964 \theta_0$). Our domain width is $L_z \simeq 117 \theta_0$, which is the same as the one by Wu et al.\ (2020).

\begin{figure}
	\begin{center}
	\includegraphics*[width=\linewidth,trim={{0.0\linewidth} {0.0\linewidth} {0.0\linewidth} {0.0\linewidth}},clip]{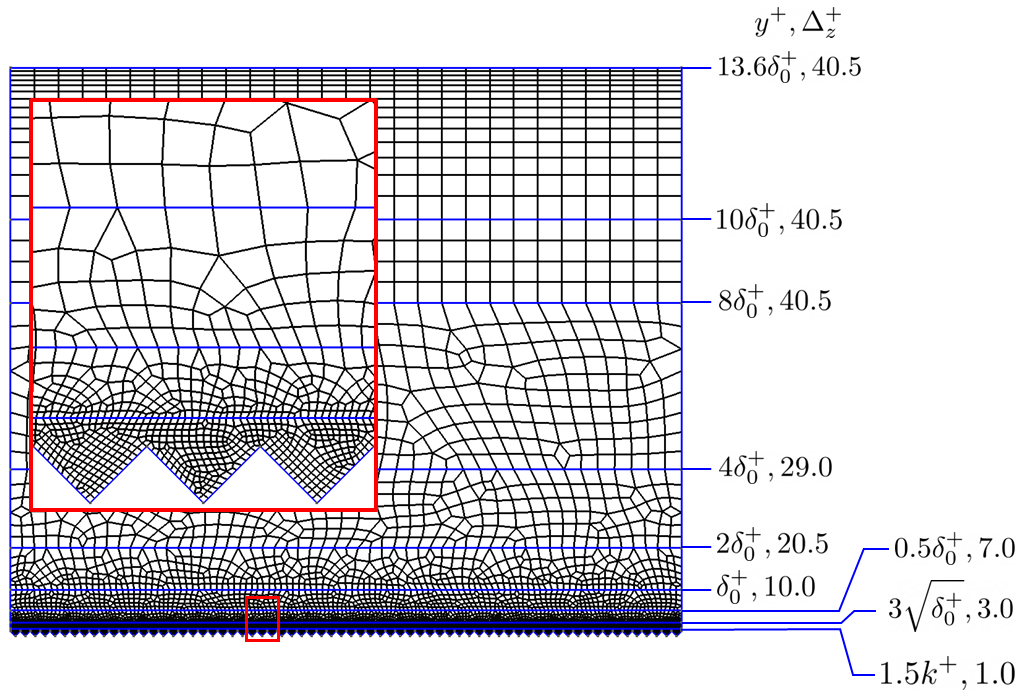}
	\caption{\label{fig:fig4} Quadratic elements on a $yz$-plane for the T9 case.}
	\end{center}
\end{figure}

\begin{table}
\centering
\begin{tabular}{cccccc}
 Case  & $ \Delta^+_x$ & $\Delta^+_\xi$ & $\Delta^+_z$ & $\Delta^+_y$   \\ \hline
 Smooth  &  $30.0$ & $-$ & $5.0 - 41.0$ & $0.25 - 40.0$  \\
 Smooth  &  $20.0$ & $-$ & $5.0 - 41.0$ & $0.25 - 40.0$ \\ 
 Smooth  &  $10.0$ & $-$ & $5.0 - 41.0$ & $0.25 - 40.0$  \\ \hline 
 T9  &  $30.0$ & $1.0$ & $0.7 - 40.5$ & $0.7 - 39.3$ \\
 T9  &  $20.0$ & $1.0$ & $0.7 - 40.5$ & $0.7 - 39.3$ \\
 T6  &  $30.0$ & $0.6$ & $0.3 - 40.5$ & $0.5 - 40.5$ \\ \hline
  \end{tabular}
\caption{Simulation cases for the separating TBL over smooth wall, as well as T9 and T6 riblets.} \label{tab:sbl}
\end{table}

We leveraged the unstructured-grid nature of SOD2D and put a considerable effort to generate an optimal grid (Figure~\ref{fig:fig4}). Our aim was to minimize the computational cost, yet achieve accurate results. The grid size distribution is inspired by the study of Chen \& He (2023). We apply a uniform grid in the streamwise direction, but we increase the grid sizes in the $yz$-plane from the bottom wall to the top boundary. We apply $\Delta^+_z \le 3.0$ up to the beginning of the logarithmic region of the ZPG TBL at $x=0$ $(y^+ \simeq 3 \sqrt{\delta^+_0})$ to well resolve the near-wall cycle of turbulence in the buffer region. Then we increase $\Delta^+_z$ to $10.0$ at the boundary layer thickness $(y^+ = \delta^+_0)$, and further increase $\Delta^+_z$ to $40.0$ by the top boundary. The wall-normal grid size follows a hyperbolic-tangent distribution from the bottom wall $(\Delta^+_y \le 0.7)$ to $y^+ = \delta^+_0 (\Delta^+_y \simeq 8.0)$. Then it further increases to $\Delta^+_y \simeq 40.0$ at $y^+ = 8\delta^+_0$. We refine $\Delta^+_y$ near the top boundary to avoid numerical instabilities due to the zero-vorticity condition. We generated this grid for all our separating TBL cases, with some differences for $y^+ \le 3 \sqrt{\delta^+_0}$ (Table \ref{tab:sbl}).

\begin{figure}
	\begin{center}
	\includegraphics*[width=\linewidth,trim={{0.0\linewidth} {0.0\linewidth} {0.0\linewidth} {0.0\linewidth}},clip]{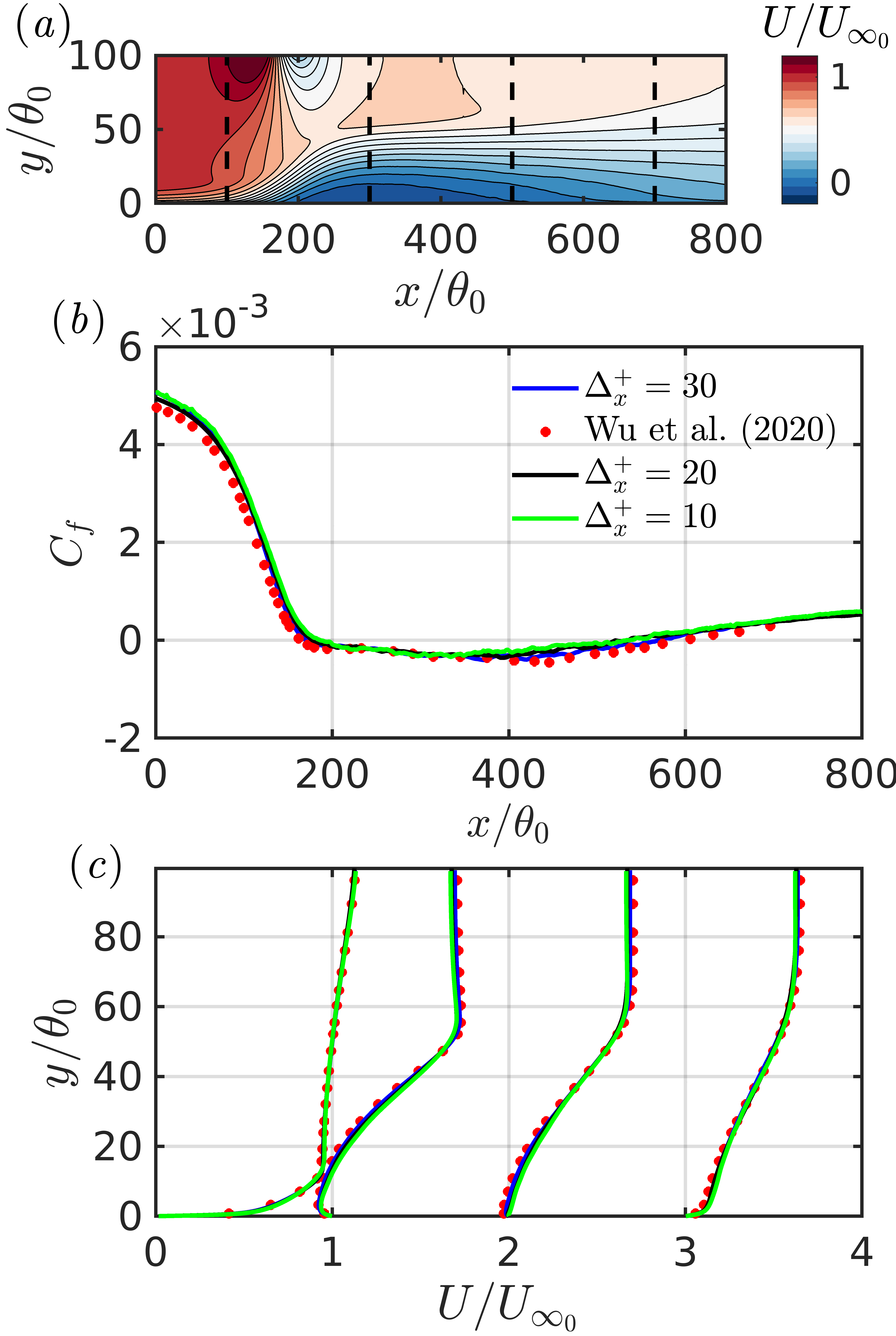}
	\caption{\label{fig:fig5} Comparison of separating TBL over smooth wall between our cases (Table \ref{tab:sbl}) and the reference study of Wu et al.\ (2020). (\textit{b}) Streamwise variation of $C_f$, and (\textit{c}) mean velocity profiles at $x/\theta_0 = 100, 300, 500, 700$ as indicated in (\textit{a}) with vertical dashed lines.}
	\end{center}
\end{figure}

We extensively validated our separating TBL setup by replicating the DNS of separating TBL over a smooth wall by Wu et al.\ (2020) (Figure~\ref{fig:fig5}), with comparable $Re_{\theta_0}$ (Figure~\ref{fig:fig3}\textit{b}), as well as matched suction profiles and location. We employed a $yz$-grid distribution as in Figure~\ref{fig:fig4}, and conducted three cases with $\Delta^+_x = 30, 20$ and $10$. Figures~\ref{fig:fig5}(\textit{b,c}) show little difference between these three cases, which are in good agreement with the data of Wu et al.\ (2020). We further refined the $yz$-grid and observed little difference in $C_f$, as well as profiles of the mean velocity $U$ and turbulent kinetic energy (TKE). We also extended the effective domain length from $L_{x_\mathrm{eff}} \simeq 991 \theta_0$ to $1390 \theta_0$ and observed no difference in $C_f, U$ and TKE. We also refined $\Delta^+_x$ from $30$ to $20$ for the T9 case, and obtained small change in $C_f$ and $U$, and only $1.5\%$ change in the mean separation location.

\begin{figure}
	\begin{center}
	\includegraphics*[width=\linewidth,trim={{0.0\linewidth} {0.0\linewidth} {0.0\linewidth} {0.0\linewidth}},clip]{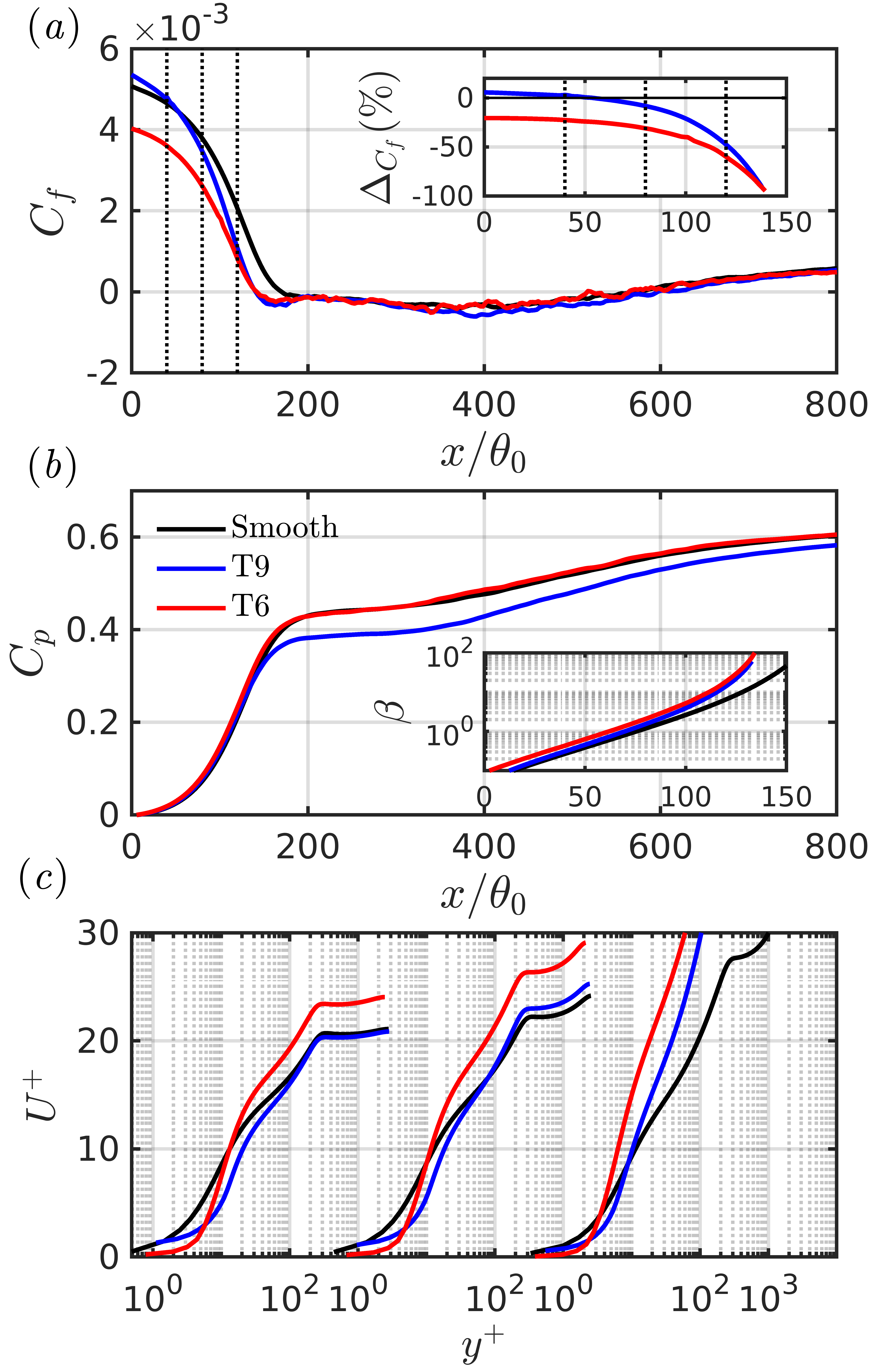}
	\caption{\label{fig:fig6} Streamwise variations of (\textit{a}) $C_f$, and (\textit{b}) wall pressure-coefficient $C_p$ and Clauser pressure-gradient parameter $\beta$. In (\textit{a}), the inset plots the percentage reduction in $C_f$ over the riblet cases compared to the smooth case. (\textit{c}) Mean velocity profiles at $x/\theta_0 = 40, 80$ and $120$, as indicated in ({\it a}) with vertical dotted lines.}
	\end{center}
\end{figure}

\section{Results}
Variations of $C_f$ indicate that both T9 and T6 cases hasten the mean separation location compared to the smooth case (Figure~\ref{fig:fig6}\textit{a}). Interestingly, the evolution of $C_f$ from $x=0$ to $x \simeq 110 \theta_0$ differs significantly between T9 and T6. However, the $C_f$ of the two riblet cases approach together from $x \simeq 110 \theta_0$ up to the mean separation location at $x \simeq 140 \theta_0$; this separation distance is $18\%$ shorter than the one for the smooth case $(x \simeq 170 \theta_0)$. For T9, $C_f$ is higher than the smooth $C_f$ at $x=0$, where there is a ZPG TBL. However, further downstream, the TBL is exposed to an APG, and the T9 $C_f$ drops below the smooth $C_f$, hence the T9 riblet reduces drag $(\Delta_{C_f} < 0)$ when is exposed to an APG (Figure~\ref{fig:fig6}\textit{a}, inset). The higher the APG, the higher the drag reduction, in accordance with previous works in the literature. The Clauser pressure-gradient parameter grows $\propto x^{2.5}$ from $\beta = 0.26$ at $x = 40\theta_0$ to $\beta = 2.73$ at $x = 100 \theta_0$ (Figure~\ref{fig:fig6}\textit{b}, inset); within this distance, $\Delta_{C_f}$ for T9 decreases from $+2.6\%$ to $-20.5\%$. 

\begin{figure}
	\begin{center}
	\includegraphics*[width=\linewidth,trim={{0.0\linewidth} {0.0\linewidth} {0.0\linewidth} {0.0\linewidth}},clip]{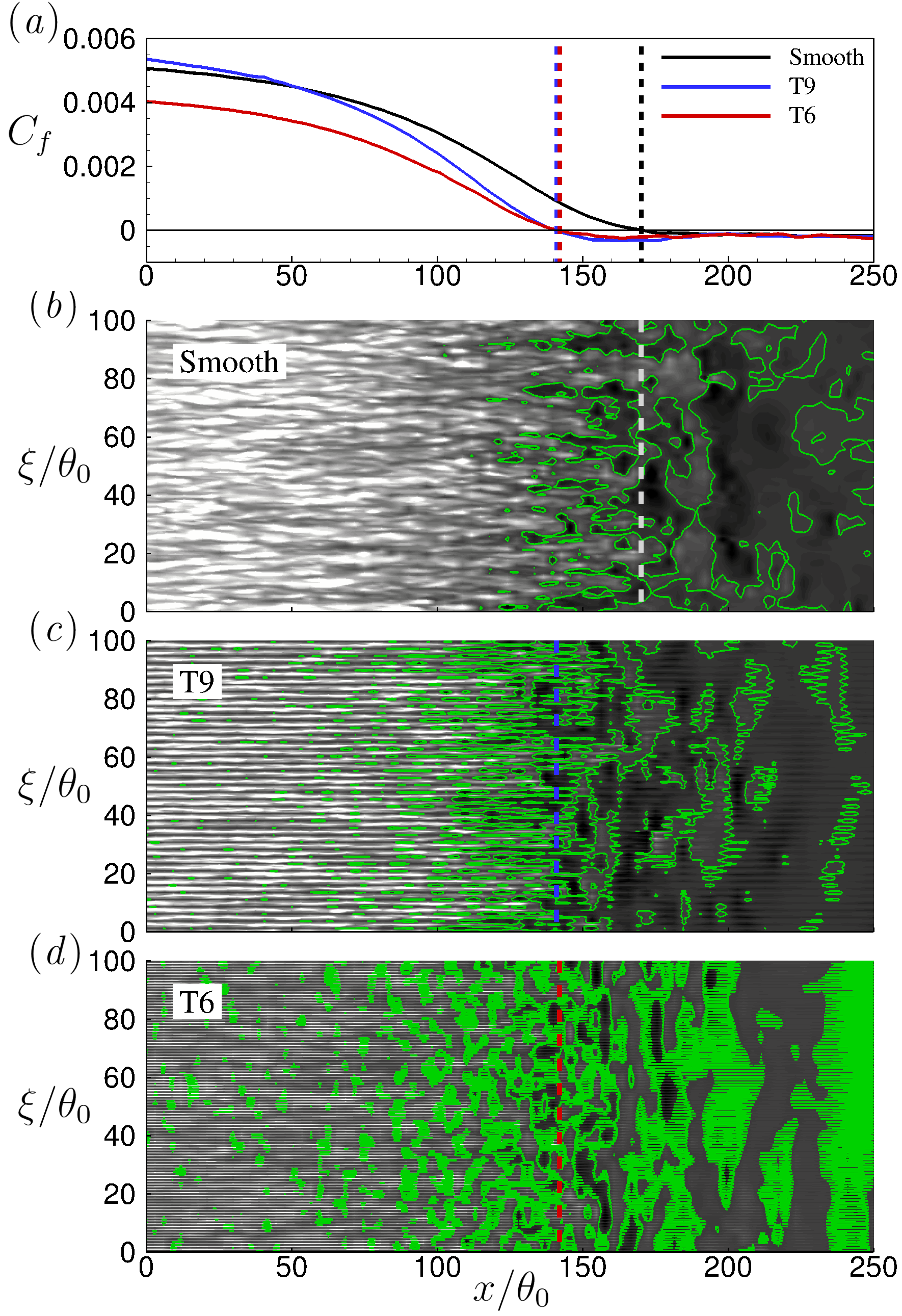}  
	\caption{\label{fig:fig7} Visualizations of the local $\hat{C}_f$ for (\textit{b}) smooth case, (\textit{c}) T9 riblets and (\textit{d}) T6 riblets; isolines of $\hat{C}_f = 0$ are highlighted in green. (\textit{a}) Streamwise variations of $C_f$, same as Figure~\ref{fig:fig6}(\textit{a}). In all panels, the mean separation location is highlighted with a vertical dashed line.}
	\end{center}
\end{figure}

The drag-reducing response of T9 under APG is also evident from the mean velocity profiles (Figure~\ref{fig:fig6}\textit{c}). At $x = 0$ ($\Delta_{C_f} = +5.6\%$), the $U^+$ profile of T9 is below the smooth case. However, at $x = 120 \theta_0$ ($\Delta_{C_f} = -47.9\%$), the $U^+$ profile of T9 falls above the smooth case for $y^+ \gtrsim 10$. For the drag-reducing T6, $\Delta_{C_f}$ becomes more negative from $x=0$ by increasing APG, similar to T9 (Figure~\ref{fig:fig6}\textit{a}, inset). However, the rate of drop in $\Delta_{C_f}$ for $0 \le x \le 120\theta_0$ is milder for T6 than T9, despite being exposed to almost equal $\beta$ (Figure~\ref{fig:fig6}\textit{b}, inset). From $x \simeq 120 \theta_0$ up to the mean separation location, $\Delta_{C_f}$ from the two riblet cases are close to each other.

The streamwise variations of $C_f$ over riblets up to the mean separation location is consistent with the study of Savino et al.\ (2025). They apply APG to an attached TBL over sinusoidal riblets with $k/s \simeq 0.95$. They consider combinations of two $\beta$ profiles ($0 \le \beta \lesssim 5$ and $0 \le \beta \lesssim 10$), and three riblet spacings ($s^+_0 = 16, 26, 60$). Their range of $\beta$ is comparable with ours for $0 \le x \lesssim 100 \theta_0$, and their lowest $s^+_0$ is comparable with our T6 case with $k/s \simeq 0.87$ and $s^+_0 \simeq 13$. For all cases, they discover that riblets under APG reduce drag as high as $100\%$. Drag reduction is higher for the riblets with larger $s^+_0$ that increase drag under ZPG.

Savino et al.\ (2025) associate the significant drag-reducing response of riblets under APG to the enhancement of the KH rollers. We conjecture that hastening the mean separation location by riblets is also related to the KH rollers. We support this conjecture by visualizing the instantaneous local $\hat{C}_f$ for the smooth case (Figure~\ref{fig:fig7}\textit{b}), compared with T9 (Figure~\ref{fig:fig7}\textit{c}) and T6 riblets (Figure~\ref{fig:fig7}\textit{d}). Similar to Figure~\ref{fig:fig1}, we plot $\hat{C}_f$ over the azimuthal $\xi$ and streamwise $x$ directions, to better identify the KH rollers. For $0 \le x \lesssim 50 \theta_0$, under mild $0 \le \beta \lesssim 0.5$, we observe sparse patches of $\hat{C}_f <0$ over T9 and T6 riblets, indicating some weak KH rollers with little coherence and large streamwise spacing. However, comparatively, the KH rollers are more coherent over the T6 riblets, consistent with Figure~\ref{fig:fig1}. From $x \simeq 50 \theta_0$ ($\beta \simeq 0.5$) up to the mean separation location ($\beta \simeq 10$), the patches of $\hat{C}_f <0$ (hence KH rollers) become larger and more coherent, with reduced spacing; this behavior is more prominent for the T6 riblets. By the mean separation location, we observe large KH rollers that are elongated over the entire domain width. These large rollers penetrate into the turbulent separation bubble. 

It is well established that KH rollers are among the prominent mechanisms for drag-increasing behavior of riblets (Garcia-Mayoral \& Jim{\' e}nez 2011, Endrikat et al.\ 2021). However, these findings consider turbulent channel flow, where KH rollers have structural differences compared to those under APG TBL. We also observe such differences between our channel flow (Figure~\ref{fig:fig1}) and separating TBL cases (Figure~\ref{fig:fig7}). In a turbulent channel flow, KH rollers (hence patches of $\hat{C}_f <0$) have a spacing $\lambda^+_x \simeq 200 - 400$ (Rowin et al.\ 2025). Patches of $\hat{C}_f <0$ are accompanied by high-shear (high $\hat{C}_f$) regions between them. On average, high $\hat{C}_f$ regions overbalance $\hat{C}_f <0$ regions, leading to $\Delta_{C_f} > 0$. Under APG, however, the spacing between patches of $\hat{C}_f <0$ is significantly reduced, leading to $\Delta_{C_f} < 0$.

\section{ Conclusions } 
We investigated the effects of riblets on the structure of TBL separation. We followed the setup of Wu et al.\ (2020), and generated a ZPG TBL with $Re_{\theta_0} = 583$ at a reference location, followed by its separation through imposition of a suction profile. We focused on two riblet geometries: 1) a triangular riblet with tip angle $\ang{90}$, that at the reference location has $s^+_0 = 52$ and increases drag, and 2) a triangular riblet with tip angle $\ang{60}$, that at the reference location has $s^+_0 = 13$ and reduces drag. We leveraged the unstructured-grid nature of our spectral-element solver, and generated an optimal grid that well resolves the riblet surface and grows in size away from the wall, consistent with the physics of TBL. We extensively validated our computational setup by replicating the DNSs of turbulent channel flow over riblets (Endrikat et al.\ 2021) and the separating TBL (Wu et al.\ 2020). Both riblet geometries shorten the mean separation distance from the reference point, $140\theta_0$ compared to $170\theta_0$ for the benchmark smooth case. Consistent with the observations of Savino et al.\ (2025), as APG increases from the reference location, both riblet geometries significantly reduce drag compared to the smooth case. Such behavior of riblets, as well as their hastening of the mean separation location is related to the enhancement of the riblet-generated KH rollers under APG. As APG increases, the KH rollers become larger with shorter streamwise spacing. By the mean separation location, we observe large coherent rollers that are elongated over the entire domain width, and travel through the turbulent separation bubble.  


\Acknowledgments
AR acknowledges the support from the Air Force Office of Scientific Research (AFOSR) under award number FA8655-24-1-7008, monitored by Dr.\ Douglas Smith and Dr.\ Barrett Flake. 
VK acknowledges his AI4S fellowship within the Generaci\'on D initiative by Red.es, Ministerio para la Transformaci\'on Digital y de la Funci\'on P\'ublica, for talent attraction (C005/24-ED CV1), funded by NextGenerationEU through PRTR.
WW acknowledges the support from the AFOSR Grant No.\ FA9550-25-1-0033, monitored by Dr.\ Gregg Abate. We thank EPSRC for the computational time made available on ARCHER2 via the UK Turbulence Consortium (EP/R029326/1), and the UKRI access to the HPC call 2024.


\begin{References}
\item Boomsma, A. and Sotiropoulos, F. (2015). Riblet drag reduction in mild adverse pressure gradients: A numerical investigation. {\it Int. J. Heat Fluid Flow}, Vol. 56, pp. 251-260.
\item Chen, C. and He, L. (2023). Two-scale solution for tripped turbulent boundary layer. {\it J. Fluid Mech.}, Vol. 955, pp. A5.
\item Choi, K.S. (1989). Near-wall structure of a turbulent boundary layer with riblets. {\it J. Fluid Mech.}, Vol. 208, pp. 417-458.
\item Choi, H., Moin, P. and Kim, J. (1993). Direct numerical simulation of turbulent flow over riblets. {\it J. Fluid Mech.}, Vol. 255, pp. 503-539.
\item Debisschop, J.R. and Nieuwstadt, F.T.M. (1996). Turbulent boundary layer in an adverse pressure gradient-effectiveness of riblets. {\it AIAA J}, Vol. 34, pp. 932-937.
\item Endrikat, S., Modesti, D., Garcia-Mayoral, R., Hutchins, N. and Chung, D. (2021). Influence of riblet shapes on the occurrence of Kelvin--Helmholtz rollers. {\it J. Fluid Mech.}, Vol. 913, pp. A37.
\item Garcia-Mayoral, R. and Jim{\' e}nez, J. (2011). Hydrodynamic stability and breakdown of the viscous regime over riblets. {\it J. Fluid Mech.}, Vol. 678, pp. 317-347.
\item Garcia-Mayoral, R., G{\'o}mez-de-Segura, G. and Fairhall, C.T. (2019). The control of near-wall turbulence through surface texturing. Fluid Dyn. Res., Vol. 51, pp. 011410.
\item Guermond, J.L., Pasquetti, R. and Popov, B. (2011). Entropy viscosity method for nonlinear conservation laws. {\it J. Comput. Phys.}, Vol. 230, pp. 4248-4267.
\item Klumpp, S., Guldner, T., Meinke, M. and Schr{\: o}der, W. (2010), Riblets in a turbulent adverse-pressure gradient boundary layer. {\it In 5th Flow Control Conference} pp. 4706.
\item Nieuwstadt, F.T.M., Wolthers, W., Leijdens, H., Krishna Prasad, K. and Schwarz-van Manen, A. (1993). The reduction of skin friction by riblets under the influence of an adverse pressure gradient. {\it Exp. Fluids}, Vol. 15, pp. 17-26.
\item Rowin, W.A., Deshpande, R., Wang, S., Kozul, M., Chung, D., Sandberg, R.D. and Hutchins, N. (2025). Experimental characterisation of Kelvin–Helmholtz rollers over riblet surfaces. {\it J. Fluid Mech.}, Vol. 1009, pp. A65.
\item Savino, B.S., Rouhi, A. and Wu, W. (2025). Attached Decelerating Turbulent Boundary Layers over Riblets. {\it arXiv preprint arXiv:2505.16962.}
\item Schlatter, P. and {\"O}rl{\"u}, R. (2010). Assessment of direct numerical simulation data of turbulent boundary layers, {\it J. Fluid Mech.}, Vol. 659, pp. 116-126.
\item Walsh, M.J. and Weinstein, L.M. (1979). Drag and heat-transfer characteristics of small longitudinally ribbed surfaces. {\it AIAA J}, Vol. 17, pp. 770-771.
\item Wu, W., Meneveau, C. and Mittal, R. (2020). Spatio-temporal dynamics of turbulent separation bubbles, {\it J. Fluid Mech.}, Vol. 883, pp. A45.
\end{References}

\end{document}